\documentstyle[osa,manuscript,psfig]{revtex}

\begin{document}

\titlepage
\title{\hskip 11.0truecm {\normalsize{FUB-HEP/96-4}}\\
{\vspace{1truecm}}
	Tracing the origin of the 
single-spin asymmetries observed in 
inclusive hadron production processes at high energies} 
\author {C. Boros, Liang Zuo-tang, Meng Ta-chung and R. Rittel}
\address {Institut f\"ur theoretische Physik,
Freie Universit\"at Berlin \\
Arnimallee 14, 14195 Berlin, Germany}
\maketitle

\begin{abstract}                

It is pointed out that the existing models for 
the left-right asymmetries observed in 
single-spin inclusive hadron production processes 
can be differentiated experimentally. 
Several such experiments are proposed with which 
the basic assumptions of these models 
can be tested individually. 

\end{abstract}     

\newpage

Striking left-right asymmetries ($A_N$) 
for hadron production
in single-spin high energy hadron-hadron collisions  
have been observed [1-9]. 
Significant effects have been seen [3-9] 
in the projectile fragmentation region in experiments 
with transversely polarized proton beams [3-9] as well as 
in those with transversely polarized anti-proton beams [5,9]. 
Data are now available not only for pions [3-6,9]
but also for $\eta $ meson [8], for kaons [9] 
and for $\Lambda$ hyperon [7].
These results are of particular interest 
for the following reasons: 
First, such experiments are conceptually simple. 
Second, the observed effects 
were unexpected theoretically [10]. 
Third, until now, little 
information has been obtained on the {\it transverse} 
spin distributions in nucleon and 
the spin-dependent hadronic interactions
in {\it transversely} polarized cases. 
Fourth, the observed asymmetries depend in particular 
on the flavor quantum numbers of the 
projectile and on those of the observed hadron [2-9]. 
Hence, they can also be used to study 
the flavor dependence of the spin distributions 
in the hadron.

A number of mechanisms [10-25] 
have been proposed recently which can give non-zero $A_N$'s 
in the framework of quantum chromodynamics (QCD) and 
quark or quark-parton models.
The existing models can approximately be divided 
into two categories: 
(1) perturbative QCD based hard scattering models [10-17] 
and (2) non-perturbative quark-fusion models [19-25]. 

In the pQCD based hard scattering models [10-17], 
the cross section for inclusive hadron production 
in hadron-hadron collision process 
is expressed as convolution of the following three factors: 
(a) the momentum distribution functions of the quarks in 
the colliding hadrons; 
(b) the cross section for the elementary hard scattering 
 between a constituent (quark, anti-quark or gluon) 
 of one of the colliding hadrons 
 with one of the other; and 
(c) the fragmentation function of the scattered 
  constituent which describes 
  its hadronization process.
The cross section for the elementary hard scattering 
can be and has been calculated [10]
using perturbative QCD. 
The obtained result [10] shows that, to the leading order, 
the asymmetry for the elementary process 
is proportional to $m_q/\sqrt{s}$ 
(where $m_q$ is the quark mass and $\sqrt{s}$ is the 
total center of mass energy 
of the colliding hadron system),  
which is negligibly small at high energies. 
In order to describe 
the observed large asymmetries in terms of 
such models, we can make use one or more 
of the following three possibilities: 
(i) Look for higher order and/or higher twist 
 effects in the elementary processes 
 which lead to larger asymmetries;  
(ii) Introduce asymmetric 
 intrinsic transverse momentum distributions 
 for the transversely polarized quarks 
 in a transversely polarized nucleon; 
(iii) Introduce asymmetric transverse momentum distributions 
 in the fragmentation functions for the transversely polarized quarks, 
 which lead to the observed hadrons.  
All three possibilities have been discussed in the literature [11-17].
We note that the question whether (or which one of) 
these possible effects indeed exist(s) is not yet settled. 
It is clear that
under the condition that perturbative QCD is indeed applicable for 
the description of such processes, 
it should (at least in principle) be possible to 
find out how significantly the effects mentioned in (i)  
contribute to $A_N$  by performing 
the necessary calculations. 
But, in contrast to this, probabilities (ii) and (iii) 
can only be probed experimentally. 
This is because 
the asymmetric momentum distributions mentioned in (ii) and (iii) 
have to be introduced by hand. 
Hence, whether 
such asymmetric distribution functions 
indeed exist,  
and how large they are if they exist,  
are questions which  
can only be answered by 
performing suitable experiments.

In the second type of models, the basic elementary 
processes are quark-antiquark fusions 
--- processes which cannot be calculated 
by using perturbative QCD. 
The fusions of the valence quarks of the 
projectile with suitable antiquarks from the target 
lead to the mesons  
observed in the projectile-fragmentation region. 
Here, the spatial extension of the colliding hadrons and thus 
hadronic surface effect plays an important role. 
Together with such surface effect, the orbital motion 
of the valence quarks in transversely polarized nucleons 
gives rise to the left-right asymmetries observed in the 
abovementioned experiments [1-9].
It has been shown [19-25] that such models 
describe the existing left-right asymmetry data[2-9]. 
In fact, they not only 
correctly predicted the change of sign 
for the asymmetry for $\pi^+$ and $\pi^-$ in reactions 
using polarized anti-proton beam [19-21], 
the non-existence of the $x_F$-$x_T$-scaling [22], 
the non-existence of asymmetry 
in the beam fragmentation region in reaction using $\pi$-beam 
and polarized nucleon target [19-21], 
but also the left-right asymmetry for $K^0$ [23]. 
Furthermore, they also predict the existence 
of left-right asymmetries for lepton-pair 
and for $K^+$ production [23,25].

Having seen that both types of models 
can [26] --- at least in principle --- 
give non-vanishing values for the left-right asymmetries 
in high energy single-spin hadron-hadron collisions, 
we ask: ``Can we locate the origin of the observed asymmetries?'' 
``Can we tell which approach 
is the more appropriate one 
by comparing such models with the 
abovementioned data [1-9] ?'' 
The answers to these questions are unfortunately ``No!''. 
The reason is not difficult to find: 
There are three possibilities, (i), (ii) and (iii), 
to obtain significant asymmetries in Type One models. 
What has been measured in such experiments [1-9] 
is the convolution of all three factors.
It is impossible to find out which one is asymmetric 
by comparing with such measurements. 
Hence, we are led to the following questions: 
Is it possible to test the 
basic assumptions of the models 
without theoretical preference or prejudice? 
Is it possible to have experiments 
with which these basic assumptions can be tested 
individually ?

In this Letter, we show that 
these questions can be answered in the 
affirmative. 
The experiments we propose are based on 
the following observations:
First, if we can determine the 
direction of motion of the polarized quark before 
it fragments into hadrons and 
measure the left-right asymmetry 
of the produced hadrons 
with respect to this direction, 
we can directly see whether the products 
of the quark hadronization process is
asymmetric with respect to this jet axis.
This means such measurements should 
be able to tell us whether 
the corresponding quark fragmentation function is asymmetric. 
Second, there is no hadronization (in other words no 
fragmentation) of the struck quarks in inclusive 
lepton-pair or $W^\pm$-production processes. 
Hence, measurements of the left-right asymmetries in 
such processes should yield useful
information on the properties of the factors other than the
quark-fragmentation function.   
Third, while surface effect may play a significant role in 
hadron-hadron collision processes, it does not 
exist in deep inelastic lepton-hadron 
scattering in large $Q^2$ and 
large $x_B$ region where 
the exchanged virtual photons 
are considered as ``bare photons'' [27]. 
Hence, comparisons between these two kinds 
of processes can yield useful information on the 
role played by surface effects. 
The experiments we propose are the following.

(A) Perform $l+p(\uparrow )\to l+\pi+X$ 
for large $x_B$ ($>0.1$, say) and large 
$Q^2$ ($>10$ GeV$^2$, say) 
and measure the left-right asymmetry 
in the current fragmentation region 
{\it with respect to the jet axis}. 
(See, in this connection, also [14]).
Here, $l$ stands for lepton which can be an $e^-$ or a $\mu^-$; 
$x_B\equiv Q^2/(2P\cdot q)$ is the usual Bjorken-$x$, 
$Q^2\equiv -q^2$, and $P, k, k', q\equiv k-k'$ are the 
four momenta of the proton, incoming electron, outgoing electron 
and the exchanged virtual photon respectively. 
The $x_B$ and $Q^2$ are chosen in 
the abovementioned kinematic region 
in order to be sure that the following is true: 
(1) The exchanged virtual photon can be treated as a bare photon [27];  
(2) This bare photon will mainly 
be absorbed by a valence quark 
which has large transverse polarization 
in a transversely polarized proton. 

Here, in this reaction, 
a valence quark is knocked out by the virtual photon
$\gamma ^*$ and fragments into 
the hadrons observed in the current jet. 
The jet direction is approximately 
the moving direction of 
the struck quark before its hadronization; 
and the struck quark has a given probability to 
be polarized transversely to this jet axis [28]. 
The transverse momenta of the produced hadrons 
with respect to this axis
come solely from the fragmentation of the quark.   
Hence, by measuring this transverse momentum distribution, 
we can directly find out whether the fragmentation function 
of this polarized quark is asymmetric.

It should also be mentioned that 
the struck quark can be a $u$ or a $d$ and  
the relative weight is $4u(x_B,Q^2):d(x_B,Q^2)$,  
which is enhanced by the corresponding charge factor.  
Taken together with the fact 
that $u$ fragments to $\pi^+$ or $\pi^0$ and 
$d$ to $\pi^-$ or $\pi^0$, we expect to see the following, 
if the left-right asymmetry indeed originates from the quark
fragmentation. The asymmetries in these processes 
should be of the same order of magnitude 
as those observed in hadron-hadron collisions.  
But $A_N(\pi^0)$ should be closer to $A_N(\pi^+)$ 
than in the hadron-hadron collision processes. 

(B) Perform the same kind of experiments as 
that mentioned in (A) and  
measure the left-right asymmetry of the produced pions 
in the current fragmentation region 
{\it with respect to the photon direction} in the 
rest frame of the proton, and examine those events 
where the lepton plane is perpendicular to the 
polarization axis of the proton.
In such events, the obtained asymmetry should 
contain the contributions 
from the intrinsic transverse motion of quarks 
in the polarized proton and  
those from the fragmentation of polarized quarks, 
provided that they indeed exist. 
That is, we expect to  
see significant asymmetries if and only if  
one of the abovementioned effects 
is indeed responsible for the asymmetries observed 
for pion production in single-spin 
hadron-hadron collisions, 
and they should be of the same order of magnitude as 
those observed in hadron-hadron collisions.  
In contrast, we should see no asymmetry if 
the observed asymmetries 
in single-spin 
hadron-hadron collisions originate from the elementary 
hard scattering or hadronic surface effects.  
Note that, 
compared with experiment (A), 
this experiment has the advantage that 
one does not need to determine the jet axis. 
But it has the disadvantage that 
one has to fix the lepton-plane 
and thus reduces the statistics of such measurements.

(C) Perform the same kind of experiments as that in (A),
but measure the left-right asymmetry 
{\it in the target fragmentation region} 
with respect to the moving direction of the proton   
in the collider (e.g. HERA) laboratory frame. 
By doing so, we are looking at the fragmentation products 
of ``the rest of the proton'' complementary to the struck quark 
(from the proton). 
Since there is no 
contribution from the elementary 
hard scattering processes and there is no hadronic  
surface effect, $A_N$ 
should be zero if the existence of left-right asymmetries is due to 
such effects. 
But, if such asymmetries originate from the fragmentation  
and/or from the intrinsic transverse motion of the 
quarks in the polarized proton,  
we should also be able to see them here. 
To be more precise, 
for $\pi^+$ and $\pi^-$ production, 
$A_N$ should be approximately 
the same as those observed 
in hadron-hadron collisions. 
But, for $\pi^0$,  
it should be less than that in the hadron-hadron case.  
This is because it is more probable 
for the virtual photon to knock out 
a $u$ valence quark from the proton.
It should also be mentioned that, 
compared to the experiment mentioned in (B), 
this experiment has the following advantages: 
Here, we do not have the spin transfer factor [14,28]
which reduces the polarization, and we do not need 
to select events according to the lepton planes.

(D) Measure the left-right asymmetry $A_N$ for $l\bar l$ and/or that for 
$W^\pm$ in $p(\uparrow )+p(0)\to l\bar l \ \mbox {or } W^\pm +X$. 
Here, if the observed $A_N$ for hadron production 
indeed originates from the quark fragmentation, we should see 
no left-right asymmetry in such processes. 
This is because we do not have any contribution 
from the quark fragmentation here.  
What we have are contributions from 
the intrinsic quark distribution functions  
including orbital motion of valence quarks 
and/or from those due to surface effect. 
We therefore expect to see significant asymmetries 
if these effects indeed exist and are responsible for 
the asymmetries observed in hadron production processes. 
The $Q^2$-dependence of the asymmetries 
reflects the $Q^2$-dependence of the 
quark distribution functions or that of the surface effect. 
(Here, $Q$ stands for the 
invariant mass of the lepton pair, 
or, in the case of $W$-production, the mass of $W$-boson). 
It is expected that the surface effect should  
depend only weakly, if at all,  on $Q^2$. 
In the figure, we show the expected $A_N$ for $l\bar l$ or $W^\pm$ 
by assuming that the surface effect does not 
depend on $Q^2$. 

Furthermore, it is useful to recall that the leading order 
elementary processes [29] to lepton-pair production 
and the first order QCD corrections [30] to them 
can be easily calculated; 
and the results are in very good agreement with 
the cross section data in the unpolarized case, 
but show zero asymmetry for single-spin hadron-hadron collisions. 
Hence, if the asymmetry indeed originates 
from the elementary hard process, 
it is expected to be very small 
for lepton-pair production.

In this paper, we have shown that 
different mechanisms which yield 
striking left-right asymmetry 
for inclusive hadron production processes
in single-spin hadron-hadron collisions 
can be differentiated experimentally. 
Several experiments have been suggested, 
with which the basic assumptions of 
such models can be checked individually.
The proposed experiments, together with the 
expected results of different models, 
are summarized in the table.
The outcomes of these experiments 
are expected to have strong impacts on the study of 
spin distribution in nucleon and on the study of spin-dependent 
hadronic interactions, in particular
in transversely polarized cases.\\

This work was supported in part by Deutsche
Forschungsgemeinschaft (DFG: Me \mbox{470/7-1)}.

\newpage 

\noindent
{\large Figure caption}
\vskip 0.6truecm
\noindent
Fig.1 
Predictions of the picture proposed in [19] on  
the left-right asymmetry $A_N$ for 
$p(\uparrow \nobreak)+p(0)\to l\bar l \mbox {\ or \ } W^\pm +X$ 
as a function of $x_F$ at $\sqrt {s}=200$ GeV. 
For lepton pair production, the solid, dashed and dotted lines are 
for $Q=4, 20 $ and 50 GeV/c$^2$ respectively.
\newpage

\widetext

\begin{table}
\caption{Predictions for left-right asymmetry in the proposed 
experiments if the asymmetry originates from the different kinds 
of effects discussed in the text.\\[-0.1cm] }
\begin{tabular}{c|c|c|c|c}
\hline 
&
\multicolumn{4}{c}{
\begin{minipage}[t]{9.6cm}
\begin{center}
\ \\[-0.3cm]
 If the $A_N$ observed in $p(\uparrow)+p(0)\to \pi +X$ 
 originates from ... 
\end{center}
\end{minipage}}\\[0.2cm]
\begin{minipage}[t]{6.3cm}
\begin{center}
  process \phantom{xxxx}
\end{center}
\end{minipage}
& 
\begin{minipage}[t]{2.2cm}
\begin{center}
quark distribution function
\end{center}
\end{minipage}
& 
\begin{minipage}[t]{2.2cm}
\begin{center}
elementary scattering process 
\end{center}
\end{minipage} 
&
\begin{minipage}[t]{2.2cm}
\begin{center}
quark fragmentation function  
\end{center}
\end{minipage}
& 
\begin{minipage}[t]{2.62cm}
\begin{center}
orbital motion \\ + \\
surface effect
\end{center}
\end{minipage} 
\\[0.9cm]  \hline 
\begin{minipage}[t]{6.3cm}
\begin{center}
\ \\[-0.1cm]
 $l+p(\uparrow)\to l+ 
 \left(\begin{array}{cc}\pi^\pm\\ K^+\\ \end{array}\right) + X $
\end{center}
\end{minipage} &  
\begin{minipage}[t]{2.2cm}
\begin{center}
\ \\[-0.3cm]
$A_N=0$ \\ wrt jet axis 
\end{center}
\end{minipage}
& 
\begin{minipage}[t]{2.2cm}
\begin{center}
\ \\[-0.3cm]
$A_N=0$ \\ wrt jet axis 
\end{center}
\end{minipage}
& 
\begin{minipage}[t]{2.2cm}
\begin{center}
\ \\[-0.3cm]
$A_N\not =0$ \\ wrt jet axis 
\end{center}
\end{minipage}
& 
\begin{minipage}[t]{2.2cm}
\begin{center}
\ \\[-0.3cm]
$A_N=0$ \\  wrt jet axis
\end{center}
\end{minipage}
\\ [-0.1cm] \cline{2-5}  
\begin{minipage}[t]{6.3cm}
\begin{center}
\ \\[-0.3cm]
 in the current fragmentation region \\ 
 for large $Q^2$ and large $x_B$ 
\end{center}
\end{minipage} &  
\begin{minipage}[t]{2.2cm}
\begin{center}
\ \\[-0.3cm]
$A_N\neq 0$ \\ wrt $\gamma^\star$ axis
\end{center}
\end{minipage}
& 
\begin{minipage}[t]{2.2cm}
\begin{center}
\ \\[-0.3cm]
$A_N=0$ \\ wrt $\gamma^\star$ axis
\end{center}
\end{minipage}
& 
\begin{minipage}[t]{2.2cm}
\begin{center}
\ \\[-0.3cm]
$A_N\not =0$ \\ wrt $\gamma^\star$ axis
\end{center}
\end{minipage}
& 
\begin{minipage}[t]{2.2cm}
\begin{center}
\ \\[-0.3cm]
$A_N=0$ \\ wrt $\gamma^\star$ axis
\end{center}
\end{minipage}
\\[0.6cm] \hline 
\begin{minipage}[t]{6.3cm}
\begin{center}
\ \\[-0.1cm]
 $l+ p(\uparrow) \to l+
 \left(\begin{array}{cc} \pi^\pm\\ K^+\\ \end{array}\right) + X $\\ 
 in the target fragmentation region \\     
 for large $Q^2$ and large $x_B$ 
\end{center}
\end{minipage} 
&  
\begin{minipage}[t]{2.2cm}
\begin{center}
\ \\[0.4cm]  $A_N\neq 0$
\end{center}
\end{minipage}
& 
\begin{minipage}[t]{2.2cm}
\begin{center}
\ \\[0.4cm] $A_N=0$ 
\end{center}
\end{minipage}
& 
\begin{minipage}[t]{2.2cm}
\begin{center}
\ \\[0.4cm] $A_N\neq 0$ 
\end{center}
\end{minipage}
& 
\begin{minipage}[t]{2.2cm}
\begin{center}
\ \\[0.4cm] $A_N=0$ 
\end{center}
\end{minipage}
\\[1.6cm] \hline 
\begin{minipage}[t]{6.3cm}
\begin{center}
\ \\[-0.1cm]
 $p + p(\uparrow)\to 
\left(\begin{array}{cc} l\bar{l}\\ W^\pm\\ \end{array}\right)+ X $\\
 in the fragmentation region of $p(\uparrow)$
\end{center}
\end{minipage} 
&  
\begin{minipage}[t]{2.2cm}
\begin{center}
\ \\[0.2cm] $A_N\neq 0$ 
\end{center}
\end{minipage}
& 
\begin{minipage}[t]{2.2cm}
\begin{center}
\ \\[0.2cm] $A_N\approx 0$ 
\end{center}
\end{minipage}
& 
\begin{minipage}[t]{2.2cm}
\begin{center}
\ \\[0.2cm] $A_N=0$ 
\end{center}
\end{minipage}
& 
\begin{minipage}[t]{2.2cm}
\begin{center}
\ \\[0.2cm] $A_N\neq 0$ 
\end{center}
\end{minipage}
\\ [1.3cm] \hline 
\end{tabular}
\end{table}


\begin{thebibliography}{mmm}
\bibitem{26} It is not clear that whether, if yes how, one can 
            describe the large left-right asymmetry 
            for $\Lambda$ production in the fragmentation region 
            by using the different pQCD based hard scattering models. 
            It seems that this data favors the 
            kind of approach proposed in [19-25].
\bibitem{28} In connection with this experiment, it is useful 
          to recall the following. A transversely polarized quark 
          in a transversely polarized nucleon remains 
          transversely polarized after it is 
          knocked out of the nucleon by an 
          unpolarized electron (or an unpolarized muon). 
          (See, e.g. [14] and the references given there.) 
          The angle between the polarization direction and the 
          lepton plane looking down stream remains the same 
          before and after the collision. 
          The magnitude of the polarization after the collision 
          is determined by that before the collision 
          times the spin transfer parameter $D$, and  
          $D=(1-y)/(1-y+y^2/2)$, 
          where $y\equiv P\cdot q/P\cdot k$. 
          We see that $D$ can be quite large for $y$ not very 
          near to one, which is usually the case 
          in the region of large $x_B$.
\end{thebibliography}
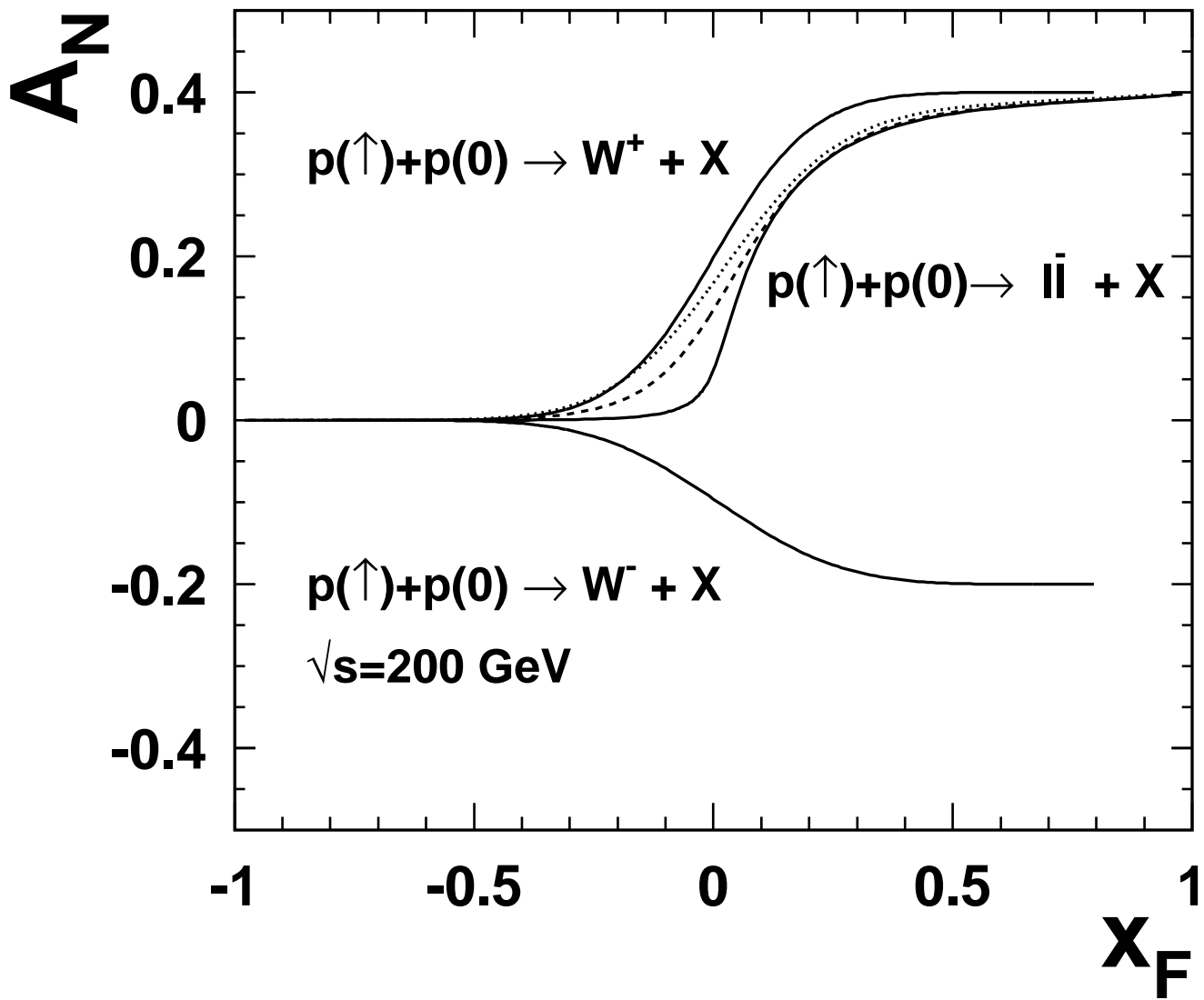
\end{document}